\documentclass[useAMS,usenatbib]{mn2e}
\usepackage{graphicx,subfigure}
\usepackage{bm}
\usepackage{longtable}
\usepackage{amsmath,amssymb,graphicx}
\usepackage{rotating}
\usepackage{color}

\newcommand{\f}[2]{\frac{#1}{#2}}

\title[Is GeV Emission from Gamma-Ray Bursts of External Shock Origin?]{Is GeV Emission from Gamma-Ray Bursts of External Shock Origin?}
\author[Amanda Maxham, Bin-Bin Zhang and Bing Zhang]{Amanda Maxham\thanks{E-mail:
maxham@physics.unlv.edu (AM); zbb@physics.unlv.edu (BBZ); zhang@physics.unlv.edu (BZ)} Bin-Bin Zhang and Bing Zhang\\
Department of Physics and Astronomy, University of Nevada, Las Vegas, Las Vegas, NV, 89124, U.S.A.}
\begin{document}

\date{Accepted . Received ; }

\pagerange{\pageref{firstpage}--\pageref{lastpage}} \pubyear{2002}

\maketitle

\label{firstpage}

\begin{abstract}
Recent observations of Gamma-Ray Bursts (GRBs) by the Fermi Large Area Telescope (LAT) revealed a power law decay feature of the high energy emission (above 100 MeV), which led to the suggestion that it originates from a (probably radiative) external shock. We analyze four GRBs (080916C, 090510, 090902B and 090926A) jointly detected by Fermi LAT and Gamma-ray Burst Monitor (GBM), which have high quality lightcurves in both instrument energy bands. Using the MeV prompt emission (GBM) data, we can record the energy output from the central engine as a function of time. Assuming a constant radiative efficiency, we are able to track energy accumulation in the external shock using our internal/external shell model code. By solving for the early evolution of both an adiabatic and a radiative blastwave, we calculate the high energy emission lightcurve in the LAT band and compare it with the observed one for each burst. The late time LAT light curves after $T_{90}$ can be well fit by the model. However, due to continuous energy injection into the blastwave during the prompt emission phase, the early external shock emission cannot account for the observed GeV flux level. The high energy emission during the prompt phase (before $T_{90}$) is most likely a superposition of a gradually enhancing external shock component and a dominant emission component that is of an internal origin.
\end{abstract}


\section{Introduction}
The Large Area Telescope (LAT; Atwood et al. 2009) aboard the Fermi Gamma Ray Space Telescope (Fermi) has recently detected nearly 20 GRBs (e.g. Abdo et al. 2009a,b; Abdo et al. 2010; Ackermann et al. 2010, see Zhang et al. 2011 for a synthetic study). Among them, several bright GRBs (e.g. GRBs 080916C, 090510, 090902B and 090926A) have well sampled long-term LAT-band lightcurves. In logarithmic space, these GRBs have count rates that rise, peak and begin decaying before the MeV prompt emission is over, i.e. peaking at a time smaller than $T_{90}$ defined in the Gamma-ray Burst Monitor (GBM; Meegan et al. 2009) detector energy band.  The post peak lightcurve typically has a decay slope steeper than $-1$ (e.g. ranging from $-1.3$ to $-2$, Ghisellini et al. 2009; Zhang et al. 2011). The simple temporal behavior (a broken power law lightcurve) of LAT emission led to the suggestion that GRB GeV emission is of an external forward shock origin (Kumar \& Barniol Duran 2009, 2010; Ghisellini et al. 2009), possibly from a highly radiative blastwave.

A simple broken power law lightcurve is expected from the blastwave evolution of an instantaneously injected fireball with fixed explosion energy. Such an approximation is valid if the analyzed time scale is much longer than $T_{90}$, the duration of the prompt gamma-ray emission itself. However, for the early blastwave evolution, especially during the epoch when the central engine is still active (as is the case for the LAT GRBs discussed in this paper), one would not expect a simple lightcurve evolution, since the energy output from the central engine is continuously injected into the blastwave.

The high quality spectral and temporal data of GRBs co-detected by Fermi LAT and GBM allow us to track the energy output from the central engine as a function of time. Recently we have developed a shell code to model the internal and external shock development for arbitrary central engine activities (Maxham \& Zhang 2009). By processing the spectral and temporal evolution data of Fermi GRBs using the method described in Zhang et al. (2011), we can model the early development of the external shock based on first hand data.

\section[]{Data Analysis}
We study four bright LAT GRBs (080916C, 090510, 090902B, and 090926A). GBM and LAT data reduction was carried out using the data analysis script introduced in Zhang et al. (2011). This code uses the public Fermi data and extracts time-resolved spectral information derived from a joint GBM/LAT fit. For the GBM data, the background spectrum is extracted using the CSPEC data, while the source spectrum is extracted using the event (TTE) data.  The LAT background is different since only a few photons are detected by LAT for most GRBs, so on-source region data long after the GBM trigger when the photon counts merge into a Poisson noise are used to derive the LAT lightcurve background.  The GBM and LAT data are then used to make dynamically time-dependent spectral fits.  The code refines the number of time slices as necessary to preserve adequate statistics in each bin, and a spectral fit is
chosen among a list of spectral models, such as a single power-law, a power-law with exponential cut-off, a Band function, a black body or a combination of these. Chi square statistics are performed to determine which fits are the best, and Ockham's Razor chooses the simplest spectral model between two statistically reasonable fits (Zhang et al. 2011).

For the 4 bright GRBs in our sample, we adopt the following models (for details, see Zhang et al. 2011). For GRB 080916C and 090926A we adopt the Band function model throughout the burst, with the spectral parameters evolving with time. GRB 090902B shows a blackbody thermal component plus a non-thermal single power law component, and the short burst GRB 090510 is best-fit with a cutoff power law plus power law component. Similar to Ghisellini et al. (2009), we found that the long-term LAT light curves decay before the end of $T_{90}$ with a slope steeper than -1.

\section[]{External shock modeling}
\subsection[]{Blastwave evolution}
We model a GRB as an explosion of many matter shells with some mass and Lorentz factor (Rees \& M\'esz\'aros 1994). As the first matter shell moves outward into the ambient medium, it slows down when sweeping up this medium (M\'esz\'aros \& Rees 1993). As time goes by, more and more trailing shells pile up onto the leading decelerating shell (Rees \& M\'esz\'aros 1998).

For an instantaneous explosion with constant energy, the motion of this decelerating ejecta along with the medium collected along the way, known as the ''blastwave", is governed by three differential equations (Chiang \& Dermer 1999; Huang et al. 2000):  radius changing with time, $\f{d R}{dt} = \beta c = \f{\sqrt{\gamma^2 - 1}}{\gamma}c$, a statement of conservation of energy and momentum across the blastwave  $\f{d \gamma}{dm} = \f{-(\gamma^2 -1)}{M}$ (Blandford \& McKee 1976), and the amount of medium swept up as a function of radius $\f{dm}{dR} = 4 \pi R^2 \rho$. Here $t$ is the time since explosion in the rest frame of the central engine, $R$ is the distance from the central engine, $\rho$ is the density of the ambient medium, $\gamma$ is the Lorentz factor of the shell, $m$ is the swept-up mass, and $M=M_0+\gamma m$ is the total effective mass including the internal energy of the blastwave, where $M_0$ is the initial mass of the ejecta. As a result, one has another differential equation,
\begin{equation}
\f{dm}{dM} = \f{1}{(1-\epsilon)\gamma + \epsilon}~,
\label{Mandm}
\end{equation}
where the value of $0\leq \epsilon \leq 1$ determines the efficiency of the radiation, with 0 representing the purely adiabatic case and 1 representing the fully radiative condition.

The above set of differential equations can be solved analytically. The adiabatic solution was presented as Eq.(14) in Maxham \& Zhang (2009). Since the LAT lightcurves decay with a slope steeper than -1 (typical value for an adiabatic blastwave), e.g. in the range of -1.3 and -2 (Zhang et al. 2011), it may be reasonable to assume a completely radiative blastwave (Ghisellini et al. 2009). By adopting a value of $\epsilon = 1$,
one can get a purely radiative solution for the blastwave, which reads
\begin{equation}
\gamma = \f{9 (M_0 \gamma_0)^2 + 12 \pi \rho M_0 R^3 (1+\gamma_0) + 8 \pi^2 \rho^2 R^6 (1+\gamma_0)
}{9 M_0^2 + 12 \pi \rho M_0 R^3 (1+\gamma_0) + 8 \pi^2 \rho^2 R^6 (1+\gamma_0)}.
\label{bwsolution}
\end{equation}
In the deceleration regime, one has $\gamma \propto R^{-3}$, and $F_\nu \propto t^{(2-6p)/7}$ for $\nu > {\rm max}(\nu_m, \nu_c)$ (which is relevant for LAT band), which is $F_\nu \propto t^{-1.6}$ for $p=2.2$ (e.g. Sari et al. 1998). This is consistent with the rapid decay observations.

\subsection{Energy injection into the blastwave}
During the prompt emission phase (i.e. $T < T_{90}$), the central engine continuously injects energy into the blastwave. So the solution should take into account the progressively increasing total energy in the blastwave. We apply the shell code developed and laid forth in Maxham \& Zhang (2009) to this problem. The code, which originally generated randomized matter shells with different mass, Lorentz factor and ejection time, is here modified to use input values for these parameters which are taken from the data as follows.

The most important parameter affecting blastwave evolution is the total injection energy. In principle the injected energy during each episode is the kinetic energy of the ejecta after energy dissipation during the prompt emission phase. Lacking a direct measure of this energy, we hereby assume that the emitted $\gamma$-ray energy is a good proxy of the kinetic energy, so that $E_k = \xi E_\gamma$. In other words, we assume a constant radiative efficiency throughout the burst. We take $\xi=1$ as the nominal value (i.e. 50\% radiative efficiency, which may be achieved for efficient magnetic energy dissipation, Zhang \& Yan 2011). In order to fit the data, we also allow $\xi > 1$ for the GRBs, which corresponds to a less efficient dissipation mechanism (e.g. in internal shocks, Panaitescu et al. 1999; Kumar 1999; Maxham \& Zhang 2009).

To evaluate $\gamma$-ray energy $E_\gamma$ as a function of time, we divide the lightcurve into multiple time bins for each burst. For each time bin (with uneven duration denoted as $\Delta T_i$ for $i$-th bin), we record its average flux $F_i$ in the GBM band, along with other useful information such as spectral parameters and the maximum photon energy.

The total gamma-ray energy released in this time bin ($i$-th) is therefore
\begin{equation}
E_{\gamma,i} = \f{4 \pi d_L^2 F_i \Delta T_i}{1 + z}.
\label{Energystart}
\end{equation}
where $z$ is the redshift (see Table 1 for values of each burst), $d_L$ is the luminosity distance of the source, and the concordance cosmology with $\Omega_{\Lambda}=0.7$ and $\Omega_m = 0.3$ is adopted in the calculation.

Adopting $E_{k,i}=\xi E_{\gamma,i}$, we then progressively increase the total energy in the blastwave $E_k = \Sigma E_{k,i}$ by adding $E_{k,i}$ in each step. For each time step, we calculate the lightcurve giving the available $E_k$. This results in a series of lightcurve solutions. The final lightcurve is then derived by jumping to progressively higher level solutions due to additional energy injections in each time step (see also Maxham \& Zhang 2009). This would result in a series of ``glitches'' in the lightcurves, each representing injection of energy from $i$-th shell into the blastwave.

Besides the energy, we also derive the (lower limit) Lorentz factor $\gamma_i$ of each shell. This parameter is important, especially for early shells, since it determines the deceleration time of a certain shell. This is particularly relevant for the first shell. The Lorentz factors of later shells are also relevant fot two reasons. First, they can be used to calculate the effective Lorentz factor of a ``merged'' shell after adding energy to an existing shell. This is needed to calculate the deceleration time of the blastwave solutions. Second, since the observed time for a late energy injection is defined by (Maxham \& Zhang 2009)
\begin{equation}
t_{\oplus,col}=t_{ej}+\frac{(t_{col}-t_{ej})}{2\gamma^2}~,
\label{collisiontime}
\end{equation}
where $t_{ej}$ and $t_{col}$ are the times of ejection and collision measured in the rest frame of the central engine. The effect of $\gamma$ becomes progressively less important, since at large $t_{ej}$'s, the second term in Eq.(\ref{collisiontime}) becomes negligible so that the observed collision time is essentially defined by the ejection time. In any case, we derive the constraints on $\gamma$ for each time bin using the pair opacity argument as described below.

To derive a constraint on the Lorentz factor, we have collected the spectral parameters and the observed maximum photon energy $E_{\rm{\oplus,max},i}$ for each time bin. One can then derive the maximum photon energy in the cosmological local frame, i.e. $E_{\rm{max},i}=E_{\rm{\oplus,max},i}(1+z)$. Requiring the pair production optical depth to be less than unity for $E = E_{\rm{max},i}$, we can write a general constraint in the parameter space of $R$ and $\gamma$ (where $R$ is the distance of the emission region from the central engine, Gupta \& Zhang 2008; Zhang \& Pe'er 2009), i.e.
\begin{equation}
R(\gamma) > \sqrt{\f{C(\beta) \sigma_T d_z^2}{-1-\beta f_0}  \left(\f{E_{\rm{max}}}{511 \rm keV^2}\right )^{-1-\beta} \left(\f{\gamma}{1+z}\right )^{2+2\beta}},
\label{Solveme}
\end{equation}
where $\sigma_T$ is the Thompson cross section, $\beta$ represents the slope of the power law component for GRBs 090902B and 090510 and the Band function high energy spectral parameter for GRBs 080916C and 090926A, and $f_0$ (in units of $\rm{ergs} \cdot \rm{cm}^{-2} \cdot s^{-1}$) can be written as $f_0 =A \cdot \Delta T \left[ \f{E_p (\alpha - \beta)}{2 + \alpha}\right]^{\alpha - \beta} \rm{exp}(\beta - \alpha)(100 \quad \rm{keV})^{-\alpha}$ for the Band function model, and $f_0=K \cdot \Delta T (100 \quad \rm{keV})^{-\beta}$ for the simple power law model, where $A$ and $K$ are normalization factors (both normalized to 100 keV). The approximation $\rm{C}(\beta) \simeq (7/6)(- \beta)^{5/3}/(1-\beta)$ (Svensson 1987) is adopted to perform the calculation.  In order to further constrain $\gamma$, one needs to make an assumption about $R$. Without other independent constraints, we apply the conventional assumption of internal shocks, so that $R(\gamma)=\gamma^2 c \f{\delta t}{1+z}$, where $\delta t$ is the observed minimum variability time scale. Combining Eq.(\ref{Solveme}), the lower limit for $\gamma$ is derived for each time bin of each burst (see also Lithwick \& Sari 2001, Abdo et al. 2009). In our calculation, we generally adopt $\gamma_i$ as the derived lower limit. This is because the derived Lorentz factors of other GRBs using the afterglow deceleration constraint (Liang et al. 2010) or photosphere constraint (Pe'er et al. 2011) are all below or consistent with these lower limits derived from the opacity constraints (Abdo et al. 2009a,b, 2010; Ackermann et al. 2010).

\subsection{Model results}

Feeding this data into our shell model code, letting each shell be ejected with energy $E_{k,i}$ and Lorentz factor $\gamma_i$ at time equal to that of the beginning of the bin time, we can calculate the early blastwave evolution and LAT band (integrated over $>100$ MeV) lightcurve for the four GRBs.

To match the observed steep decay (with slope $\sim -1.5$), we adopt a radiative fireball solution or an adiabatic fireball solution with steep electron energy index. Even though each solution (for a fixed kinetic energy) has a steep decay slope, the overall lightcurve shows a shallower decay due to piling up of successive shells ejected later, with glitches introduced by jumping among the solutions. As an example, the radiative model lightcurve of GRB 080916C as compared with observation is presented in Fig.1. The top panel shows the long term evolution, while the bottom panel is the zoomed-in early afterglow lightcurve. The dotted lines denote the blastwave solutions with progressively increasing total energy. The lowest one corresponds to the first time bin, the second lowest corresponds to adding the energy of the second time bin, etc.

Since the lightcurve is chopped into discrete time bins, the blastwave energy is added in discrete steps. This introduces some artificial glitches in the lightcurve. Such an approximation is more realistic for GRBs with distinct emission episodes. For GRB 080916C, the lightcurve is more appropriately approximated as a continuous wind with variable luminosity. The artificial glitches should appear to be more smeared. For this reason, we have smoothed the glitches to make more natural transitions between solutions. The model afterglow parameters (the fraction of electron energy $\epsilon_e$, the fraction of magnetic energy $\epsilon_B$, and the number density $n$) are presented in Table 1. These are in general consistent with the parameter constraints derived by Kumar \& Barniol Duran (2009, 2010).

In general, the model lightcurve of GRB 080916C cannot fit the early LAT data. Making the model suitable to fit the late-time steep decay, the early model lightcurve level is too low to account for the observed data. Alternatively, one can make the early model lightcurve match the observed flux level. Then inevitably the late time afterglow level exceeds the observed level significantly due to the continuous energy injection. We believe that if the LAT band emission after $T_{90}$ originates from the external shock, then the LAT emission during the prompt emission phase {\em cannot} be solely interpreted by the external shock model. The external shock contribution is relatively small, especially during early epochs when energy in the blastwave is small. As a result, the GeV emission during the prompt phase must be of an internal origin. This is consistent with the fact that the entire GBM/LAT emission during the prompt phase can be well fit by a single Band-function spectral model in all the time bins (Abdo et al. 2009a; Zhang et al. 2011).

We have also modeled GRBs 090510, 090902B and 090926A. The model parameters (for both radiative and adiabatic solutions) are listed in Table 1, and the results for radiative solution are shown in Fig.\ref{fluxlightcurves}.  In all cases, the slope and flux level of the data are matched in the latter part of the curve only. During the prompt emission phase, the data points rise above the flux prediction of the external shock model, suggesting that GeV emission is a superposition of external and internal components during the prompt emission phase ($T<T_{90}$). This conclusion is valid for both the adiabatic and radiative solutions. The difference between the two is that the adiabatic model invokes a shallower $p$ but a larger $\xi$ (and hence a larger energy budget) to fit the same data.

\begin{table}
\caption{Parameters used for the four sample bursts\label{tbl}}
\begin{tabular}{|c|c|c|}
\hline
 080916C
 & adiabatic & radiative \\
 \hline
 $p$ & 2.5 & 2.1 \\
 $\xi$ & 5 & 10\\
 $\epsilon_e$ & 0.3 & 0.3\\
 $\epsilon_B$ & 0.01 & 0.01\\
 $n$ & 1 & 1 \\
 \hline
 $z$ & 4.35\footnotemark[1] \\
 \hline
\end{tabular}

\begin{tabular}{|c|c|c|}
\hline
 090510
 & adiabatic & radiative \\
 \hline
 $p$ & 2.4 & 2.1 \\
 $\xi$ & 2 & 3\\
 $\epsilon_e$ & 0.5 & 0.5\\
 $\epsilon_B$ & 0.01 & 0.01\\
 $n$ & 0.1 & 0.1 \\
 \hline
 $z$ & 0.903\footnotemark[2]\\
 \hline
\end{tabular}

\begin{tabular}{|c|c|c|}
\hline
 090902B
 & adiabatic & radiative \\
 \hline
 $p$ & 2.4 & 2.1 \\
 $\xi$ & 1 & 1\\
 $\epsilon_e$ & 0.2 & 0.15\\
 $\epsilon_B$ & 0.001 & 0.01\\
 $n$ & 0.001 & 0.01 \\
 \hline
 $z$ & 1.822\footnotemark[3] \\
 \hline
\end{tabular}

\begin{tabular}{|c|c|c|}
\hline
 090926A
 & adiabatic & radiative \\
 \hline
 $p$ & 2.4 & 2.2 \\
 $\xi$ & 3 & 3\\
 $\epsilon_e$ & 0.3 & 0.3\\
 $\epsilon_B$ & 0.01 & 0.01\\
 $n$ & 0.1 & 0.1 \\
 \hline
 $z$ & 2.1062\footnotemark[4] \\
 \hline
\end{tabular}
\footnotetext[1]{van der Horst\& Goldstein 2008}
\footnotetext[2]{Rau et al. 2009}
\footnotetext[3]{Cucchiara et al. 2009}
\footnotetext[4]{Malesani et al. 2009}
\end{table}

\begin{figure}
\includegraphics[scale=0.8]{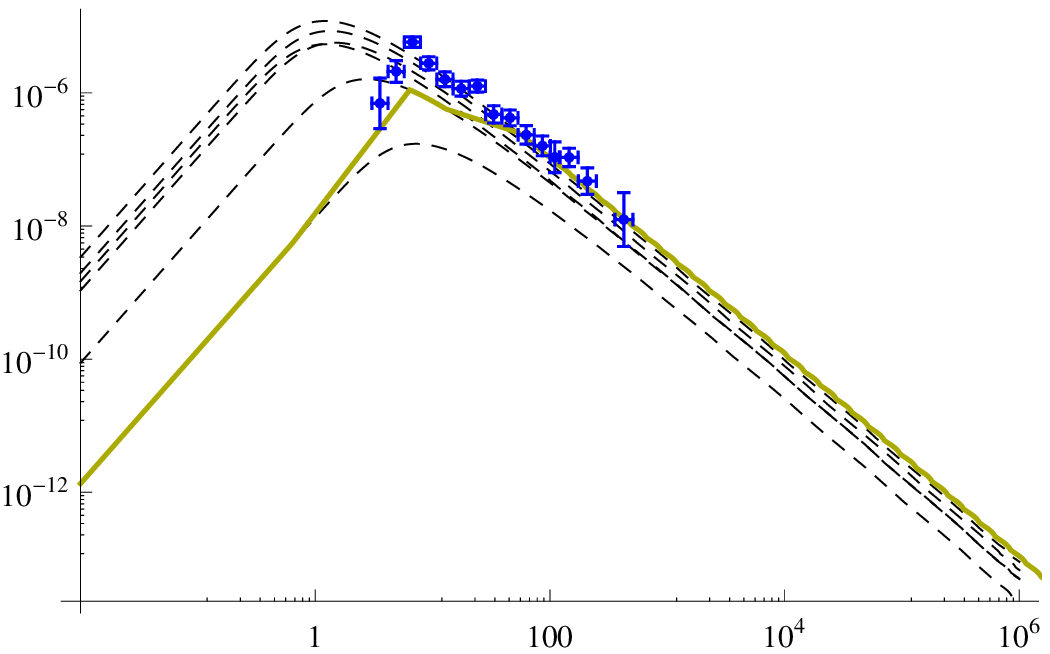}
\includegraphics[scale=0.8]{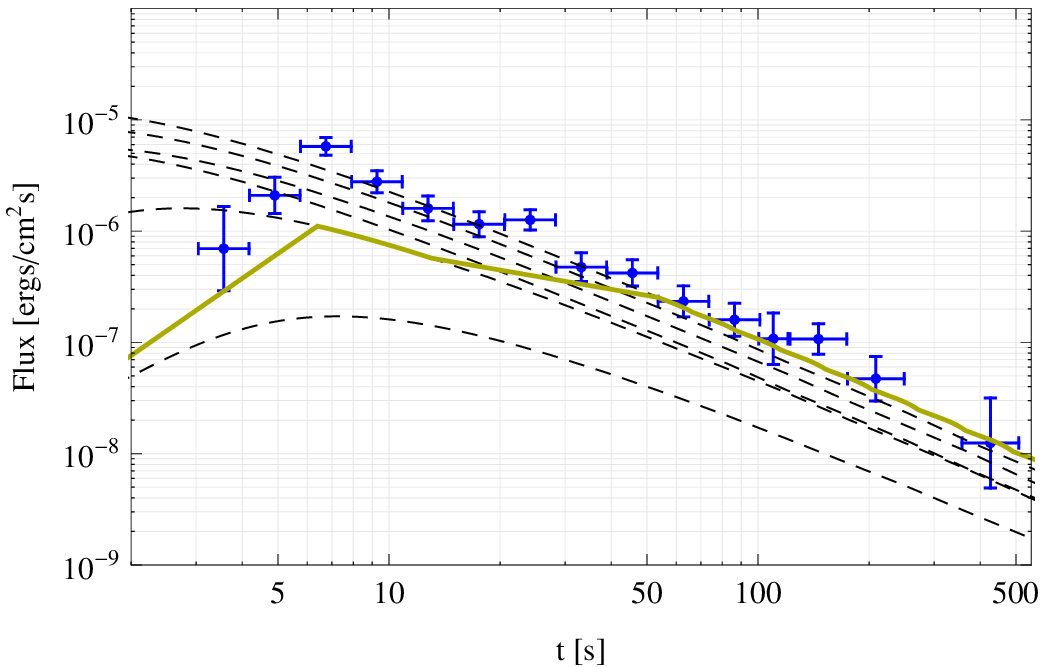}
\caption{The predicted external shock $>100$ MeV lightcurve of GRB 080916C for a radiative blastwave solution (yellow line) as compared with the data (blue points).  Successive lightcurves that correspond to different total blastwave kinetic energy are shown as dashed lines.  The top panel shows the global lightcurve, while the bottom panel shows a zoom view where the flux deficit at early times can be clearly seen.}
\label{chopbw}
\end{figure}

\begin{figure}
\includegraphics[scale=0.8]{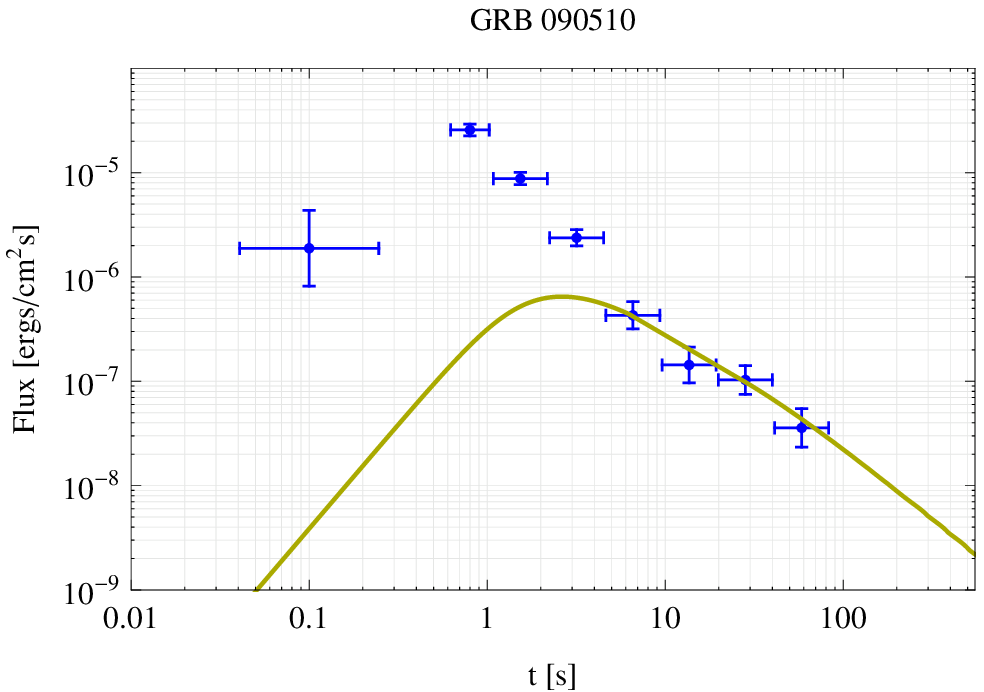}
\includegraphics[scale=0.8]{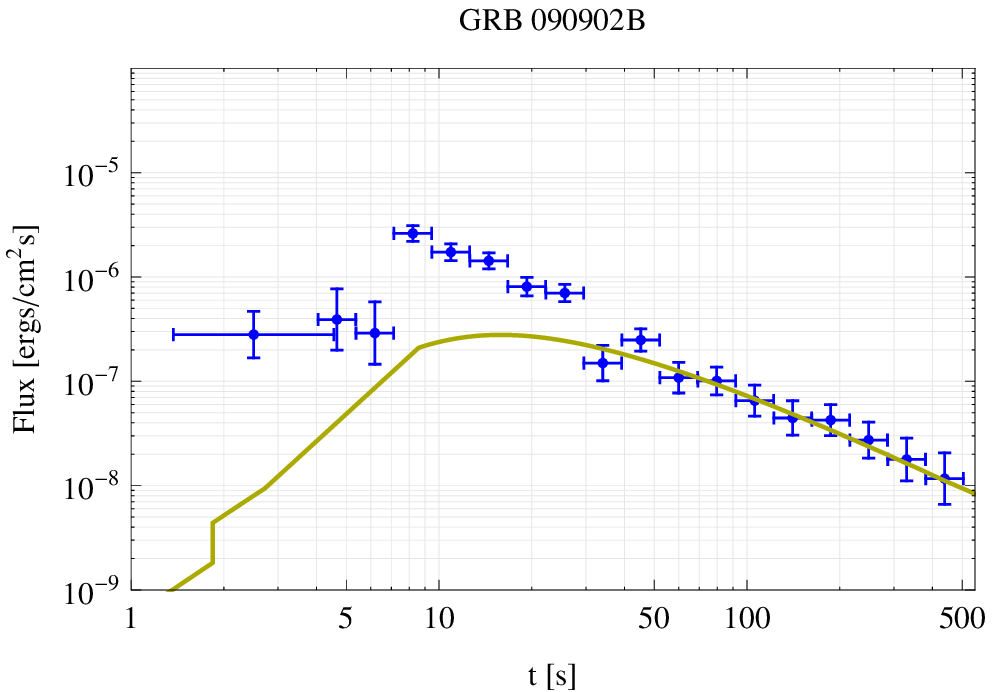}
\includegraphics[scale=0.8]{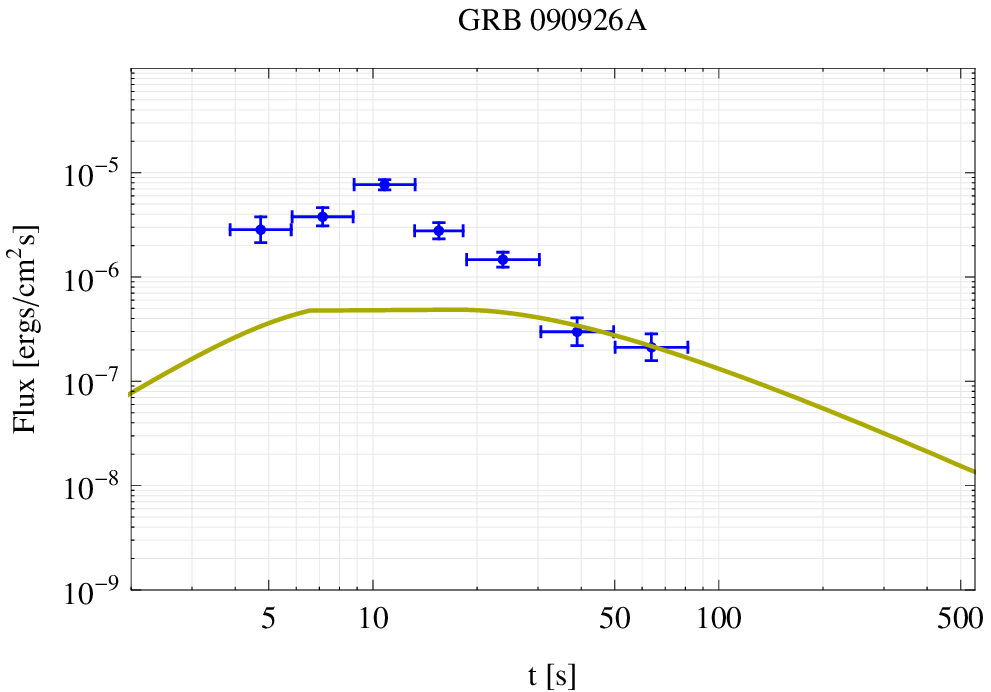}
\caption{Model predictions of $>100$ MeV lightcurve (for a radiative blastwave solution) vs. observed data for GRBs 090510, 090902B, and 090926A. The conventions are similar to Fig.1, but without successive solutions specifically plotted.}
\label{fluxlightcurves}
\end{figure}

\section{Conclusions and Discussion}

Using the first-hand Fermi data, we have tracked the energy output from the central engine and modeled the early blastwave evolution of four bright LAT GRBs. The predicted $>100$ MeV lightcurve is found unable to account for the observed LAT emission during the prompt emission phase. The main reason is that during the phase when the central engine is still active, the forward shock is continuously refreshed by late energy injection, so that the afterglow decays much slower than the case predicted by an instantaneously ejected constant energy fireball. This suggests that at least during the prompt emission phase, the LAT band emission is not of external forward shock origin. This is in contrast to the suggestion of Ghisellini et al. (2009), Kumar \& Barniol Duran (2009) and Feng \& Dai (2010), who did not consider the energy accumulation during the prompt emission phase and interpreted the entire GeV emission as due to the external shock origin. Our conclusion is based on the assumption that GRB radiative efficiency is essentially a constant throughout the burst. In order to interpret the entire afterglow as due to the external forward shock origin, one needs to ``artificially" assume that the GRB efficiency increases with time, so that the late time central engine activity, even though producing bright $\gamma$-ray emission, adds little kinetic energy into the blastwave. We believe that such an assumption is contrived.

Our conclusion is consistent with some independent arguments. From data analysis, Zhang et al. (2011) showed that during the prompt emission phase the GeV emission and MeV emission traces each other well. For GRB 080916C, the entire GBM/LAT emission can be modeled by a single Band function component in all the time bins (see also Abdo et al. 2009a). For GRB 090902B, even though GeV emission belongs to a distinct spectral component, its flux seems to track the flux of the MeV component nicely, suggesting a connection in the physical origin (see Pe'er et al. 2011 for modeling). A more definite argument in favor of an internal origin of GeV emission in GRB 080916C is that the GeV lightcurve peak coincides the second peak in the GBM lightcurve, suggesting that GeV emission is the spectral extension of MeV emission to higher energies (Zhang et al. 2011). Also individual case studies of GRB 090902B (Pe'er et al. 2011; Liu \& Wang 2011) and GRB 090510 (He et al. 2011) all suggest that the external shock model cannot interpret the prompt GeV data. In general, our modeling suggests that it is possible to use the external shock model to interpret GeV emission after the prompt emission phase, but not during the prompt emission phase (see also Kumar \& Barniol Duran 2010).

Our conclusion also has implications for understanding GRB prompt emission physics, in particular, the composition of the GRB outflow. The internal origin of GeV emission in GRB 080916C makes it essentially impossible to interpret the entire Band spectrum with the photosphere model (e.g. Beloborodov 2010; Lazzati \& Begelman 2010). The lack of photosphere emission then demands a Poynting-flux-dominated outflow at least for this burst (Zhang \& Pe'er 2009; Fan 2010), and new models in the Poynting flux dominated regime (e.g. Zhang \& Yan 2011) are called for.

\smallskip

This work is partially supported by NSF AST-0908362 and NASA NNX09AT66G, NNX10AD48G. We thank Xue-Feng Wu for helpful discussion.

\bsp

\label{lastpage}

\end{document}